%% file: phep_2015_085-arxiv.tex
\documentclass[manyauthors,nocleardouble,COMPASS,american]{cernphprep}

\usepackage{graphicx}
\usepackage[dvipsnames,usenames]{color}
\usepackage{wrapfig}
\usepackage{amsmath,amssymb}
\usepackage{eurosym}
\usepackage{bm}       
\usepackage[english]{babel}

\usepackage{amssymb}
\usepackage{amsfonts}

\usepackage{upgreek}

\input myunits.sty

\usepackage{fix2col}

\usepackage[font=small,format=plain,labelfont=bf,up]{caption}

\usepackage[T1]{fontenc}
\usepackage{mathrsfs}
\usepackage{rotating}
\usepackage[utf8]{inputenc}
\usepackage{dsfont}
\usepackage{cite}
\usepackage[dvipsnames]{xcolor}
\usepackage{pdfpages}
\usepackage{multirow}
\usepackage{array}
\usepackage{subcaption}
\usepackage{hyperref}
\usepackage{pgfplots}
\usepackage[percent]{overpic}

\newboolean{bw_mode} 
\setboolean{bw_mode}{false} 
\ifthenelse{
		\boolean{bw_mode}
	}{
		\graphicspath{{bw_gfx/}{gfx/}}
	}{
		\graphicspath{{gfx/}}
	}

\RequirePackage[T1]{fontenc}

\makeatletter
\g@addto@macro\bfseries{\boldmath}
\makeatother

\begin{document}
\selectlanguage{english}

\begin{titlepage}

\PHnumber{2015--085}
\PHdate{30 March 2015}

\title{ The  Spin Structure Function \boldmath{$g_1^{\rm p}$} of the Proton 
        and a Test of the Bjorken Sum Rule}

\Collaboration{The COMPASS Collaboration}
\ShortAuthor{The COMPASS Collaboration}
\begin{abstract}
New results for the double spin asymmetry $A_1^{\rm p}$ and the proton
longitudinal spin structure function $g_1^{\rm p}$ are presented. They were
obtained by the COMPASS collaboration using polarised 200 GeV muons scattered
off a longitudinally polarised NH$_3$ target. The data were collected in 2011
and complement those recorded in 2007 at 160\,GeV, in particular at lower values
of $x$. They improve the statistical precision of $g_1^{\rm p}(x)$ by about a
factor of two in the region $x\lesssim 0.02$.  A next-to-leading order QCD fit
to the $g_1$ world data is performed. It leads to a new determination of the
quark spin contribution to the nucleon spin, $\Delta \Sigma$ ranging from 0.26
to 0.36, and to a re-evaluation of the first moment of $g_1^{\rm p}$.  The
uncertainty of $\Delta \Sigma$ is mostly due to the large uncertainty in the
present determinations of the gluon helicity distribution.  A new evaluation of
the Bjorken sum rule based on the COMPASS results for the non-singlet structure
function $g_1^{\rm NS}(x,Q^2)$ yields as ratio of the axial and vector coupling
constants $|g_{\rm A}/g_{\rm V}| = 1.22 \pm 0.05~({\rm stat.}) \pm 0.10~({\rm
  syst.})$, which validates the sum rule to an accuracy of about 9\%.
\end{abstract}

\vspace*{60pt}
Keywords: COMPASS; deep inelastic scattering; spin; structure function; QCD analysis; 
parton helicity distributions; Bjorken sum rule.
\vspace*{60pt}
\begin{flushleft}
\end{flushleft}
\vfill
\Submitted{(to be submitted to Phys. Lett. B)}
\end{titlepage}
{\pagestyle{empty} \input{Authors2015_g1.tex}}
\newpage
\section{Introduction}
The determination of the longitudinal spin structure of the nucleon became one
of the important issues in particle physics after the surprising EMC result that
the quark contribution to the nucleon spin is very small or even vanishing
\cite{emc}.  The present knowledge on the longitudinal spin structure function
of the proton, $g_1^{\rm p}$, originates from measurements of the asymmetry
$A_1^{\rm p}$ in polarised lepton nucleon scattering. In all these experiments,
longitudinally polarised high-energy leptons were scattered off longitudinally
polarised nucleon or nuclear targets. At SLAC and JLab electron beams were used,
electron and positron beams at DESY and muon beams at CERN. Details on the
performance of these experiments and a collection of their results can be found
e.g. in Ref.\,\citen{aidala}.

In this Letter, we report on new results from the COMPASS experiment at CERN. By
measuring $A_1^{\rm p}$, we obtain results on $g_1^{\rm p}$ in the deep
inelastic scattering (DIS) region. They cover the range from $1\,(\GeV/{c})^2$
to $190\,(\GeV/{c})^2$ in the photon virtuality $Q^2$ and from 0.0025 to 0.7 in
the Bjorken scaling variable $x$.  The new data, which were collected in 2011 at
a beam energy of $200\,\GeV$, complement earlier data taken in 2007 at
$160\,\GeV$ that covered the range $0.004 < x < 0.7$ \cite{g1p2010}. In the
newly explored low-$x$ region, our results significantly improve the statistical
precision of $g_1^{\rm p}$ and thereby allow us to decrease the low-$x$
extrapolation uncertainty in the determination of first moments.

In the following section, the COMPASS experiment is briefly described. The data
selection procedure is presented in Section~3 and the method of asymmetry
calculation in Section~4. The results on $A_1^{\rm p}(x,Q^2)$ and $g_1^{\rm
  p}(x,Q^2)$ are given in Section~5. A new next-to-leading order (NLO) QCD fit
to the existing nucleon $g_1$ data in the region $Q^2 > 1\,(\GeV/{c})^2$ is
described in Section~6. Section~7 deals with the determination of first moments
of $g_1^{\rm p}$ and the evaluation of the Bjorken sum rule using COMPASS data
only. Conclusions are given in Section~8.

\section{Experimental setup}
The measurements were performed with the COMPASS setup at the M2 beam line of
the CERN SPS. The data presented in this Letter correspond to an integrated
luminosity of $0.52\,$fb$^{-1}$.  A beam of positive muons was used with an
intensity of $10^7\,{\rm s}^{-1}$ in a $10\,\s$ long spill every $40\,\s$. The
nominal beam momentum was 200~GeV/$c$ with a spread of 5\%. The beam was
naturally polarised with an average polarisation $P_{\rm B}=0.83$, which is
known with a precision of $0.04$.  Momentum and trajectory of each incoming
particle were measured in a set of scintillator hodoscopes, scintillating fibre
and silicon detectors.  The beam was impinging on a solid-state ammonia (NH$_3$)
target that provides longitudinally polarised protons.  The three protons in
ammonia were polarised up to $|P_{\rm T}| \approx 0.9$ by dynamic nuclear
polarisation with microwaves. For this purpose, the target was placed inside a
large-aperture superconducting solenoid with a field of $2.5\,\T$ and cooled to
$60\,\mK$ by a mixture of liquid $^3$He and $^4$He. The target material was
contained in three cylindrical cells with a diameter of $4\,\Cm$, which had
their axes along the beam line and were separated by a distance of $5\,\Cm$. The
outer cells with a length of $30\,\Cm$ were oppositely polarised to the central
one, which was $60\,\Cm$ long. In order to compensate for acceptance differences
between the cells, the polarisation was regularly reversed by rotation of the
magnetic field direction. In order to guard against unknown systematic effects,
once during the data taking period the direction of the polarisation relative to
the magnetic field was reversed by exchanging the microwave frequencies applied
to the cells. Ten NMR coils surrounding the target material allowed for a
measurement of $P_{\rm T}$ with a precision of 0.032 for both signs of the
polarisation. The typical dilution due to unpolarisable material in the target
amounts to about 0.15.

The experimental setup allowed for the measurement of scattered muons and
produced hadrons. These particles were detected in a two-stage, open forward
spectrometer with large acceptance in momentum and angle.  Each spectrometer
stage consisted of a dipole magnet surrounded by tracking detectors.
Scintillating fibre detectors and pixel GEM detectors in the beam region were
supplemented with Micromegas and GEM detectors close to the beam and MWPCs,
drift chambers and straw detectors that covered the large outer areas. Scattered
muons were identified in sets of drift-tube planes located behind iron and
concrete absorbers in the first and second stages.  Particle identification with
the RICH detector or calorimeters is not used in this measurement. The
`inclusive triggers' were based on a combination of hodoscope signals for the
scattered muons, while for `semi-inclusive' triggers an energy deposit of hadron
tracks in one of the calorimeters was required, optionally in coincidence with
an inclusive trigger. A detailed description of the experimental setup can be
found in Ref.\,\citen{nimpaper}.

\section{Data selection}
The selected events are required to contain a reconstructed incoming muon, a
scattered muon and an interaction vertex.  The measured incident muon momentum
has to be in the range $185\,\GeV/c < p_{\rm B} < 215\,\GeV/c$. In order to
equalise the beam flux through all target cells, the extrapolated beam track is
required to pass all of them.  The measured longitudinal position of the vertex
allows us to identify the target cell in which the scattering occurred. The
radial distance of the vertex from the beam axis is required to be less than
$1.9\,\Cm$, by which the contribution of unpolarised material is minimised.  All
physics triggers, inclusive and semi-inclusive ones, are included in this
analysis.  In order to be attributed to the scattered muon, a track is required
to pass more than $30$ radiation lengths of material and it has to point to the
hodoscopes that have triggered the event.  In order to select the region of deep
inelastic scattering, only events with photon virtuality $Q^2 > 1\,(\GeV/c)^2$
are selected. In addition, the relative muon energy transfer, $y$, is required
to be between $0.1$ and $0.9$.  Here, the lower limit removes events that are
difficult to reconstruct, while the upper limit removes the region that is
dominated by radiative events. These kinematic constraints lead to the range
$0.0025 < x < 0.7$ and to a minimum mass squared of the hadronic final state,
$W^2$, of $12\,(\GeV/c^2)^2$.  After all selections, the final sample consists
of $77$ million events. The selected sample is dominated by inclusive triggers
that contribute $84\%$ to the total number of triggers.  The semi-inclusive
triggers mainly contribute to the high-$x$ region, where they amount to about
half of the triggers. In the high-$Q^2$ region the semi-inclusive triggers
dominate.

\section{Asymmetry calculation}
\label{sec:asym}
The asymmetry between the cross sections for antiparallel\,($\uparrow
\downarrow$) and parallel\,($\uparrow \uparrow$) orientations of the
longitudinal spins of incoming muon and target proton is written as
\begin{equation}
A^{\rm p}_{\rm LL} = \frac                              
{\sigma^{\uparrow \downarrow} - \sigma^{\uparrow \uparrow}}
{\sigma^{\uparrow \downarrow} + \sigma^{\uparrow \uparrow}}\,.
\end{equation}
This asymmetry is related to the longitudinal and transverse spin asymmetries
$A_1^{\rm p}$ and $A_2^{\rm p}$, respectively, for virtual-photon absorption by
the proton:
\begin{equation}
 A^{\rm p}_{\rm LL} = D (A_1^{\rm p} + \eta A_2^{\rm p})~.
\end{equation}
The factors 
\begin{equation}
\eta = \frac{\gamma(1-y-\gamma^2y^2/4-y^2m^2/Q^2)}{(1+\gamma^2y/2)(1-y/2)-y^2m^2/Q^2}
\end{equation}
and
\begin{equation}
  D = \frac{y ( (1+\gamma^2 y/2)(2-y) - 2y^2 m^2/Q^2 )}{ y^2 (1-2m^2/Q^2)
    (1+\gamma^2) + 2(1+R) (1-y-\gamma^2 y^2/4)} 
\end{equation}
depend on the event kinematics, with $\gamma = 2 M x /\sqrt{Q^2}$.  The
virtual-photon depolarisation factor $D$ depends also on the ratio $R =
\sigma_{\rm L}/\sigma_{\rm T}$, where $\sigma_{\rm L}$ ($\sigma_{\rm T}$) is the
cross section for the absorption of a longitudinally (transversely) polarised
virtual photon by a proton.  The asymmetry $A_1^{\rm p}$ is defined as
\begin{equation}
A_1^{\rm p} = \frac{\sigma_{1/2} - \sigma_{3/2}}{ \sigma_{1/2} + \sigma_{3/2}}~,
\end{equation}
where $\sigma_{1/2} (\sigma_{3/2})$ is the absorption cross section of a
transversely polarised virtual photon by a proton with total spin projection
$\frac{1}{2} \left(\frac{3}{2}\right)$ in the photon direction. Since both
$\eta$ and $A_2^{\rm p}$ \cite{e155_a2} are small in the COMPASS kinematic
region, $A^{\rm p}_1 \simeq A^{\rm p}_{\rm LL}/D~$ and the longitudinal spin
structure function is given by
\begin{equation}
g_1^{\rm p} = \frac{F_2^{\rm p}}{2x~(1 + R)} A_1^{\rm p},
\label{extr_g1}
\end{equation}  
where $F_2^{\rm p}$ denotes the spin-independent structure function of the
proton.

The number of events, $N_i$, collected from each target cell before and after
reversal of the target polarisation is related to the spin-independent cross
section $\overline{\sigma}=\sigma_{1/2} + \sigma_{3/2}$ and to the asymmetry
$A_1^{\rm p}$ as
\begin{equation}
N_{i} = a_{i} \phi_{i} n_{i} {\overline \sigma} (1 + P_{\rm B} P_{\rm T} f D
A_1^{\rm p})~, ~~~ {i=o1,c1,o2,c2}~. 
\label{asym}
\end{equation}
Here, $a_{i}$ is the acceptance, $\phi_{i}$ the incoming muon flux, $n_{i}$ the
number of target nucleons and $f$ the dilution factor, while $P_{\rm B}$ and
$P_{\rm T}$ were already introduced in Section 2.  Events from the outer target
cell are summed, thus the four relations of Eq.~(\ref{asym}) corresponding to
the two sets of target cells (outer, $o$ and central, $c$) and the two spin
orientations (1 and 2) result in a second-order equation in $A_1^{\rm p}$ for
the ratio $(N_{o1} N_{c2})/(N_{c1} N_{o2})$. Fluxes and acceptances cancel in
this equation, if the ratio of acceptances for the two sets of cells is the same
before and after the magnetic field rotation \cite{long_pp}.  In order to
minimise the statistical uncertainty, all quantities used in the asymmetry
calculation are evaluated event by event with the weight factor \cite{long_pp}
\begin{equation}
w = P_{\rm B} f D \,.
\end{equation}
The polarisation of the incoming muons as a function of the beam momentum is
obtained from a parametrisation based on a Monte Carlo simulation of the beam
line.  The effective dilution factor $f$ is given by the ratio of the total
cross section for muons on polarisable protons to the one on all nuclei in the
target, whereby their measured composition is taken into account.  It is
modified by a correction factor that accounts for the dilution due to radiative
events on unpolarised protons \cite{terad}.  The target polarisation is not
included in the event weight, because it may change in time and generate false
asymmetries.  The obtained asymmetries are corrected for spin-dependent
radiative effects according to Ref.\,\citen{polrad} and for the $^{14}$N
polarisation as described in Refs.\,\citen{g1p2010,rondon}.  It has been checked
that the use of semi-inclusive triggers does not bias the determination of
$A_1^{\rm p}$.

Systematic uncertainties are calculated taking into account multiplicative and
additive contributions to $A_1^{\rm p}$.  Multiplicative contributions originate
from the uncertainties of the target polarisation, the beam polarisation, the
dilution factor (mainly due to the uncertainty of $R$) and the depolarisation
factor.  When added in quadrature, these uncertainties result in a total
uncertainty $\Delta A_1^{\rm mult}$ of $0.07 A_1^{\rm p}$. They are shown in
Table~\ref{tab:syst_unc}, which also shows the additive contributions.  The
largest additive contribution to the systematic uncertainty is the one from
possible false asymmetries.  Its size is estimated with two different
approaches. In the first approach, the central target cell is artificially
divided into two consecutive $30\,\Cm$ long parts. Combining these two $30\,\Cm$
long targets with the outer ones with the same polarisation, two independent
false asymmetries are formed. Both are found to be consistent with zero.  In
order to check for time-dependent effects, in the second approach the data
sample is divided into sub-samples each consisting of periods of stable data
taking with both field directions for each target cell. The results for
$A_1^{\rm p}$ obtained from these sub-samples are compared by using the method
of ``pulls'' \cite{compass_npb765}. No significant broadening of pull
distributions is observed. These pulls are used to set an upper limit on the
systematic uncertainty due to false asymmetries $A_1^{\rm false}$. Depending on
the $x$-bin, values between 0.4$\cdot\sigma_{\rm stat}$ and
0.84$\cdot\sigma_{\rm stat}$ are obtained. Further additive corrections
originate from neglecting $A_2$ and from the uncertainty in the correction
$A_1^{\rm RC}$ to the asymmetry $A_1$, which is due to spin-dependent radiative
effects.

\begin{table}
\begin{center}
\caption{Contributions to the systematic uncertainty on $A_1^{\rm p}$ with
  multiplicative (top) and additive (bottom) components.} 
\label{tab:syst_unc}
\begin{tabular}{|c|c|c|}
\hline
Beam polarisation  		& $\Delta P_{\rm B} / P_{\rm B}$ & $ 5\% $ \\
Target polarisation 		& $\Delta P_{\rm T} / P_{\rm T}$ & $ 3.5\%$ \\
Depolarisation factor 	        & $\Delta D(R)/D(R)$ & $ 2.0\; -\; 3.0 \%$ \\
Dilution factor 		& $\Delta f/f$	     & $2\% $ \\
Total 				& $\Delta A_1^{\rm mult}$ & $\simeq 0.07 A_1 $ \\
\hline
\hline
False asymmetry 		& $A_{\rm false}$	     & $<0.84\cdot \sigma_{\rm stat}$ \\
Transverse asymmetry 	        & $\eta \cdot A_2$   & $ < 10^{-2}$ \\
Radiative corrections 		& $A_1^{\rm RC}$  & $ 10^{-4}\;-\;10^{-3}$ \\
\hline
\end{tabular}
\end{center}
\end{table}
  
\section{Results on $A_1^{\rm p}$ and $g_1^{\rm p}$}
\label{sec:results}
\begin{figure}
	\centering
	\includegraphics[width=0.9\textwidth]{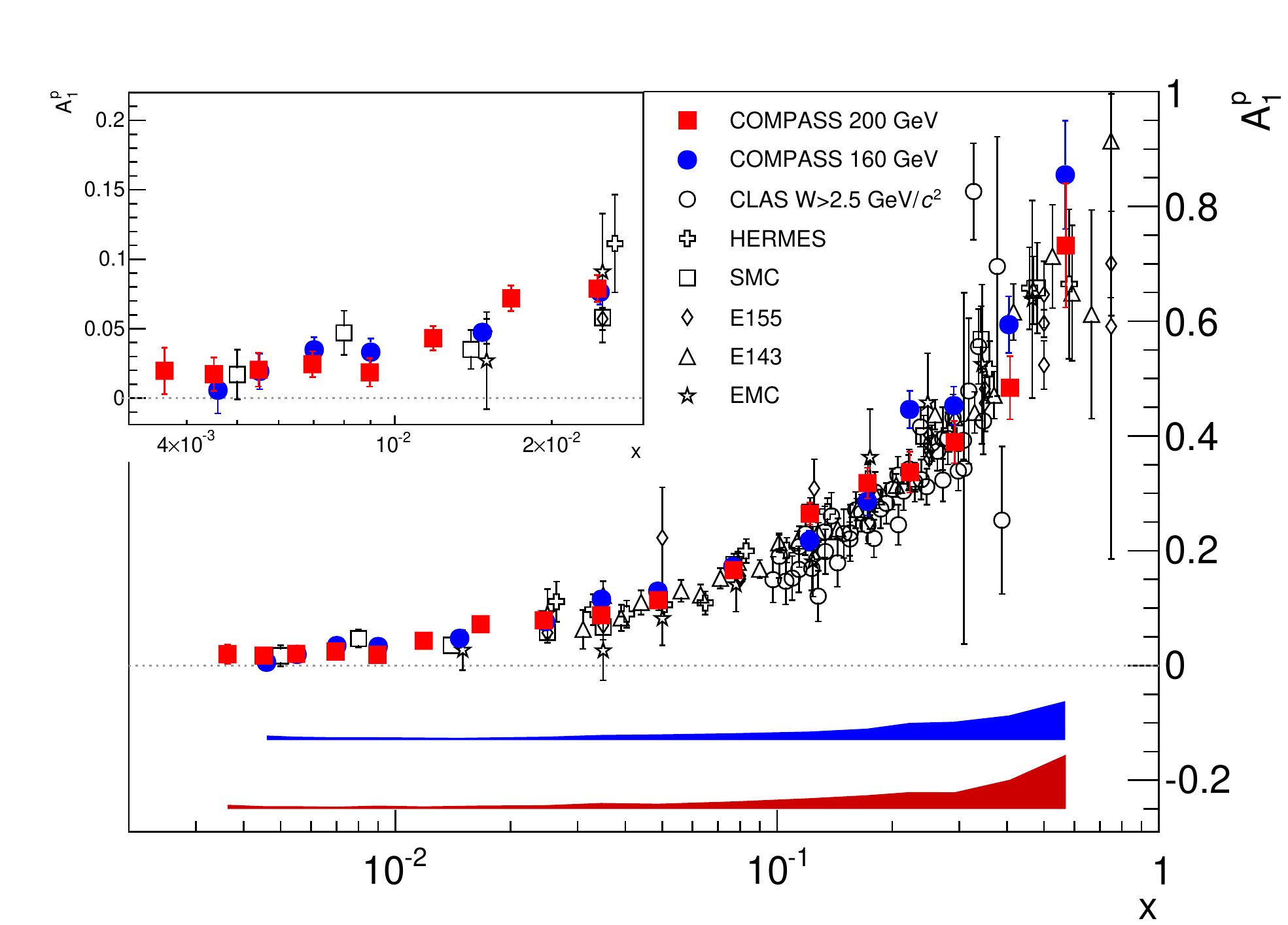}
	\caption{The asymmetry $A_1^{\rm p}$ as a function of $x$ at the
          measured values of $Q^2$ as obtained from the COMPASS data at
          $200\,\GeV$. The new data are compared to the COMPASS results obtained
          at $160\,\GeV$ \cite{g1p2010} and to the other world data (EMC
          \cite{emc}, CLAS \cite{CLAS}, HERMES \cite{HERMES}, E143 \cite{e143},
          E155 \cite{e155p}, SMC \cite{smc}).  The bands at the bottom indicate
          the systematic uncertainties of the COMPASS data at $160\,\GeV$ (upper
          band) and $200\,\GeV$ (lower band).}
\label{fig:A1_world}
\end{figure}

\begin{figure}
	\centering
	\includegraphics[width=0.9\textwidth]{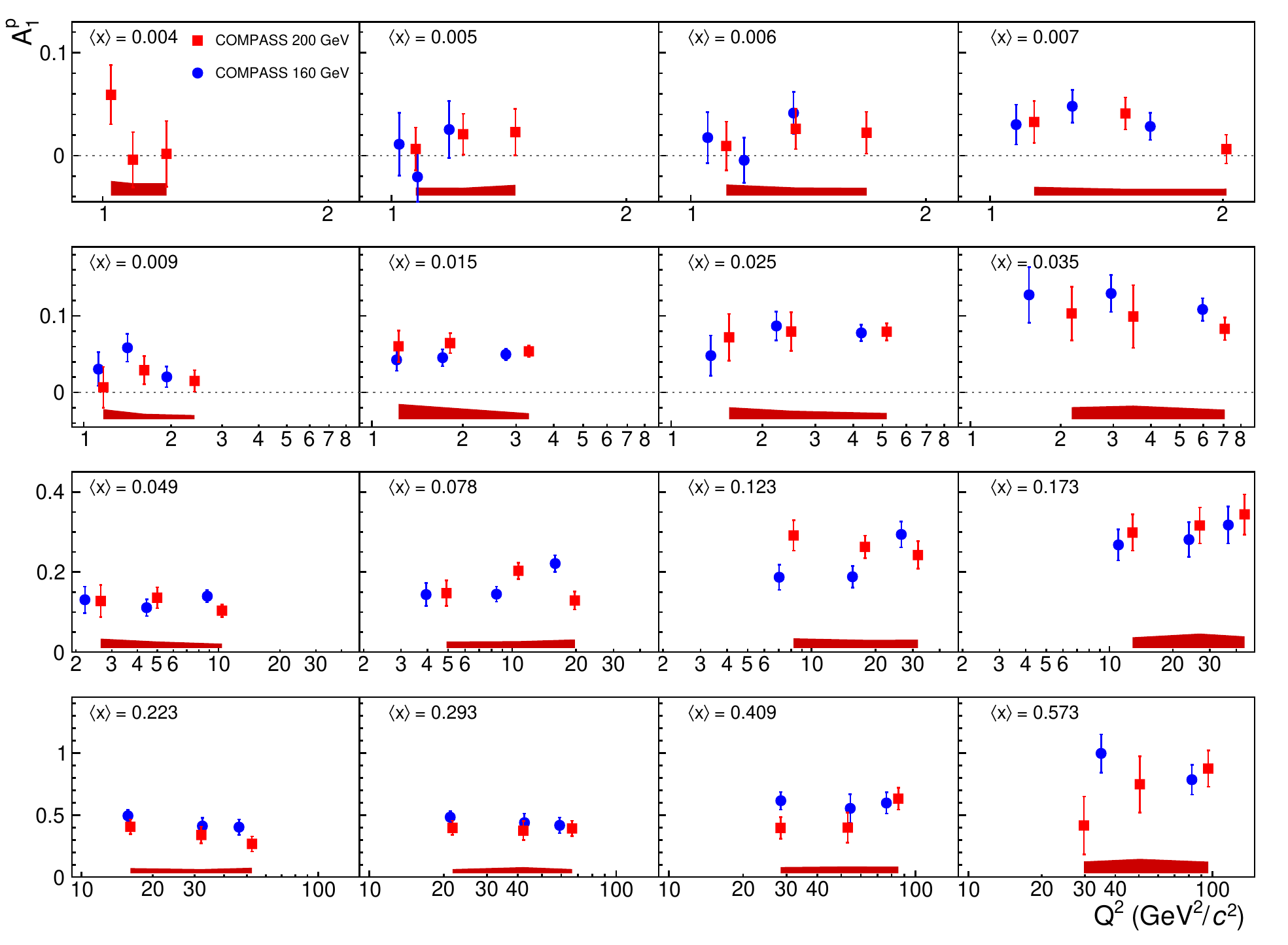}
	\caption{The asymmetry $A_1^{\rm p}$ as a function of $Q^2$ in bins of
          $x$ obtained from the $200\,\GeV$ (red squares) and $160\,\GeV$ (blue
          circles) COMPASS data. The band at the bottom indicates the systematic
          uncertainty for the $200\,\GeV$ data.}
\label{fig:A1p_x_q2}
\end{figure}

The data are analysed in terms of $A_1$ and $g_1$ as a function of $x$ and
$Q^2$.  The $x$ dependence of $A_1^{\rm p}$ averaged over $Q^2$ in each $x$ bin
is shown in Fig.\,\ref{fig:A1_world} together with the previous COMPASS results
obtained at $160\,\GeV$~\cite{g1p2010} and with results from other
experiments~\cite{emc,CLAS,HERMES,e143,e155p} including those by SMC at
$190\,\GeV$\cite{smc}. The bands at the bottom represent the systematic
uncertainties of the COMPASS results as discussed in Section~\ref{sec:asym}. The
new data improve the statistical precision at least by a factor of two in the
low-$x$ region, which is covered by the SMC and COMPASS measurements only.  The
good agreement between all experimental results reflects the weak $Q^2$
dependence of $A_1^{\rm p}$.  This is also illustrated in
Fig.\,\ref{fig:A1p_x_q2}, which shows $A_1^{\rm p}$ as a function of $Q^2$ in
sixteen intervals of $x$ for the COMPASS data sets at $160\,\GeV$ and
$200\,\GeV$. In none of the $x$ bins, a significant $Q^2$ dependence is
observed.  The numerical values of $A_1^{\rm p}(x, Q^2)$ obtained at $200\,\GeV$
are given in the Appendix.

\begin{figure}
	\centering
	\includegraphics[width=0.7\textwidth]{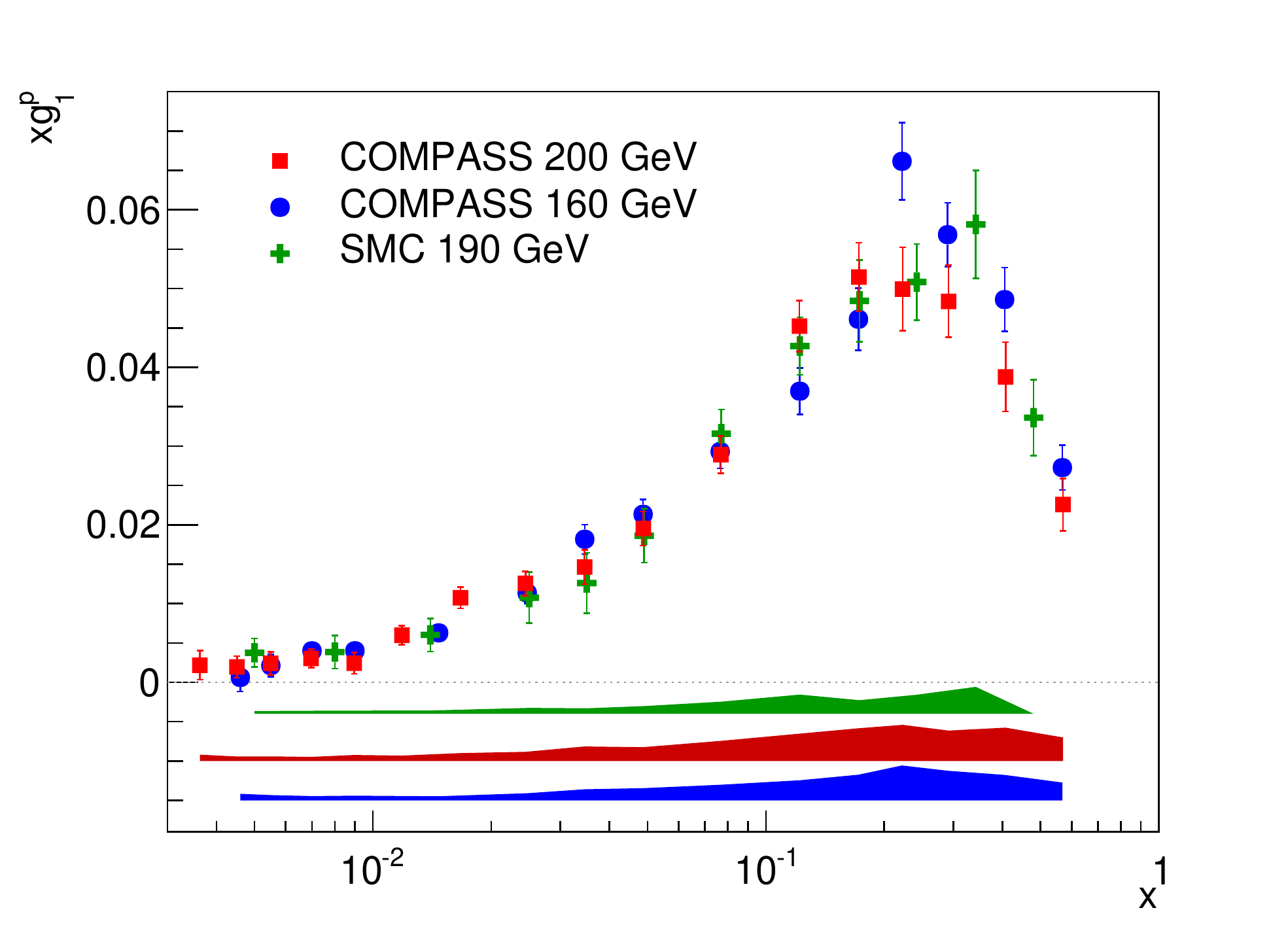}
	\caption{The spin-dependent structure function $xg_1^{\rm p}$ at the
          measured values of $Q^2$ as a function of $x$. The COMPASS data at
          $200\,\GeV$ (red squares) are compared to the results at $160\,\GeV$
          (blue circles) and to the SMC results at $190\,\GeV$ (green crosses)
          for $Q^2>1 \text{~(GeV/}c)^2$.  The bands from top to bottom indicate
          the systematic uncertainties for SMC $190\,\GeV$, COMPASS $200\,\GeV$
          and COMPASS $160\,\GeV$.  }
\label{fig:g1}
\end{figure}

The longitudinal spin structure function $g_1^{\rm p}$ is calculated from
$A_1^{\rm p}$ using Eq.~(\ref{extr_g1}), the $F_2^{\rm p}$ parametrisation from
Ref.\,\citen{smc} and the ratio $R$ from Ref.\,\citen{e143_R}.  The new results
are shown in Fig.\,\ref{fig:g1} at the measured values of $Q^2$ in comparison
with the previous COMPASS results obtained at $160\,\GeV$ and with SMC results
at $190\,\GeV$.  Compared to the SMC experiment, the present systematic
uncertainties are larger due to a more realistic estimate of false asymmetries,
which is based on real events.

\begin{figure}
	\centering
	\includegraphics[width=0.9\textwidth]{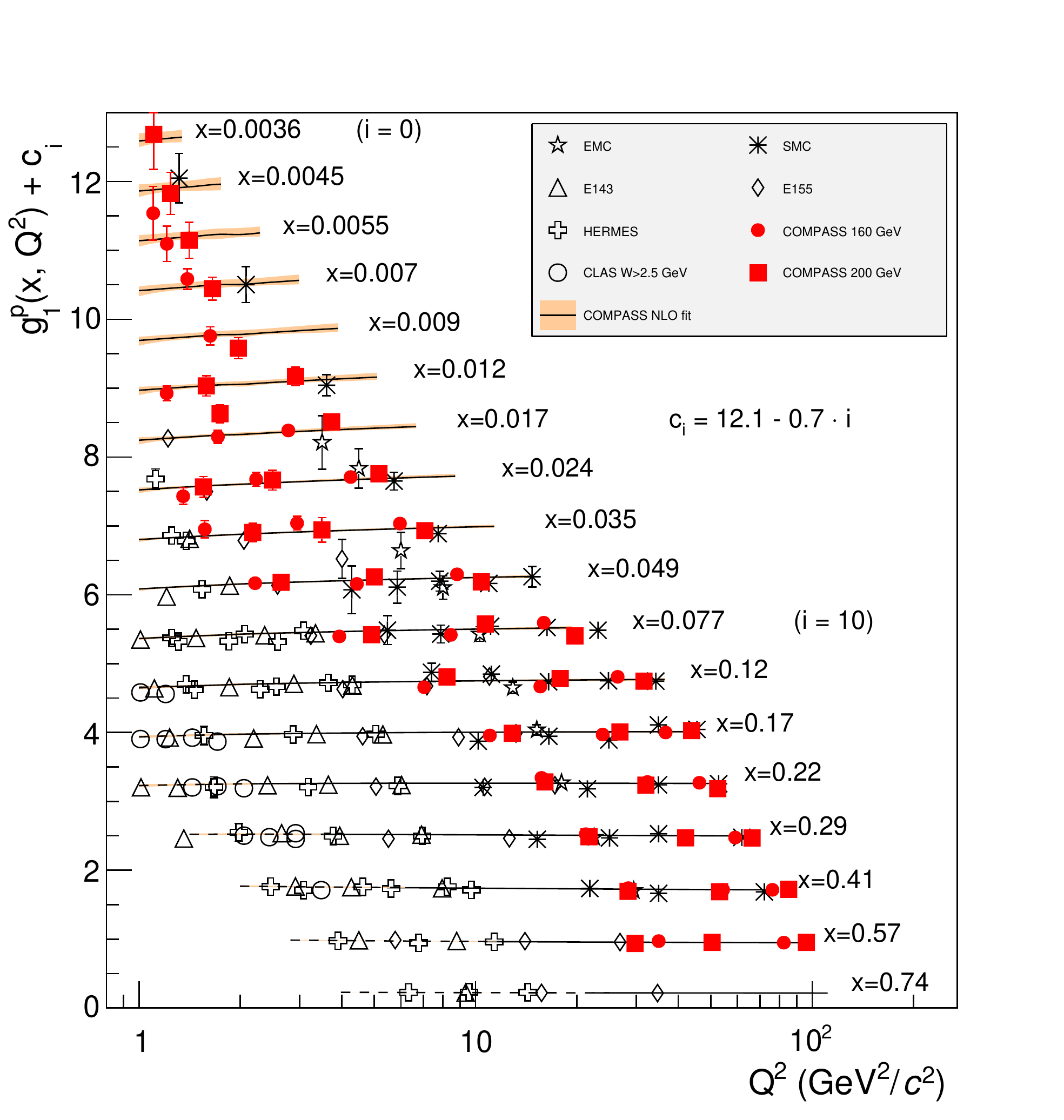}
	\caption{World data on the spin-dependent structure function $g_1^{\rm
            p}$ as a function of $Q^2$ for various values of $x$ with all
          COMPASS data in red (full circles: $160\,\GeV$, full squares:
          $200\,\GeV$). The lines represent the $Q^2$ dependence for each value
          of $x$, as determined from a NLO QCD fit (see
          Section~\ref{sec:QCD}). The dashed ranges represent the region with
          $W^2 < 10 ~(\GeV/c^2)^2$.}
\label{fig:g1p_world_data}
\end{figure}

The world data on $g_1^{\rm p}$ as a function of $Q^2$ for various $x$ are shown
in Fig.\,\ref{fig:g1p_world_data}.  The data cover about two decades in $x$ and
in $Q^2$ for most of the $x$ range, except for $x<0.02$, where the $Q^2$ range
is much more limited.  The new data improve the kinematic coverage in the region
of high $Q^2$ and low $x$ values, which gives a better lever arm for the
determination of quark and gluon polarisations from the DGLAP evolution
equations. In addition, the extension of measurements to lower values of $x$ is
important to better constrain the value of the first moment of $g_1^{\rm p}$.

\section{NLO QCD fit of $g_1$ world data}
\label{sec:QCD}
We performed a new NLO QCD fit of the spin-dependent structure function $g_1$ in
the DIS region, $Q^2~>~1\,(\GeV/c)^2$, considering all available proton,
deuteron and $^3$He data.  The fit is performed in the $\overline{\rm MS}$
renormalisation and factorisation scheme. For the fit, the same program is used
as in Ref.\,\citen{g1d2006}, which was derived from program 2 in
Ref.\,\citen{smc}.  The region $W^2 < 10~(\GeV/c^2)^2$ is excluded as it was in
recent analyses \cite{nocera}.  Note that the impact of higher-twist effects
when using a smaller $W^2$ cut is considered in Ref.\,\citen{jam}.  The total
number of data points used in the fit is 495 (see Table~\ref{tab:DataSetInput}),
the number of COMPASS data points is 138.

The neutron structure function $g_1^{\rm n}$ is extracted from the $^3$He data,
while the nucleon structure function $g_1^{\rm N}$ is obtained as
\begin{equation}
g_1^{\rm N}(x,Q^2) = \frac{1}{1-1.5~\omega_{\rm D}} g_1^{\rm d}(x,Q^2),
\label{neutron}
\end{equation}
where $\omega_{\rm D}$ is a correction for the D-wave state in the deuteron,
$\omega_{\rm D} = 0.05 \pm 0.01$ \cite{machleidt}, and the deuteron structure
function $g_1^{\rm d}$ is given per nucleon.  The quark singlet distribution
$\Delta q^{\rm S}(x)$, the quark non-singlet distributions $\Delta q_3(x)$ and
$\Delta q_8(x)$, as well as the gluon helicity distribution $\Delta g(x)$, which
appear in the NLO expressions for $g_1^{\rm p}$, $g_1^{\rm n}$ and $g_1^{\rm N}$
(see e.g. Ref.\,\citen{smc}), are parametrised at a reference scale $Q_0^2$ as
follows:
\begin{equation}
\Delta f_k (x) = \eta_k ~\frac{ x^{\alpha_k} \,(1-x)^{\beta_k} \,(1+\gamma_k x
  ~)}{\int_{0}^{1} x^{\alpha_k} \,(1-x)^{\beta_k} \,(1+\gamma_k x ~) \text{d}x}
\,.
\label{eq:param_fk}
\end{equation}
Here, $\Delta f_k (x)$ ($k=1...4$) represents $\Delta q^{\rm S}(x)$, $\Delta
q_{3}(x)$, $\Delta q_8(x)$ and $\Delta g(x)$ and $\eta_k$ is the first moment of
$ \Delta f_k(x)$ at the reference scale. The moments of $\Delta q_3$ and $\Delta
q_8$ are fixed at any scale by the baryon decay constants ($\rm F$$+$$\rm D$)
and ($\rm3F$$-$$\rm D$), respectively, assuming SU(2)$_{\rm f}$ and SU(3)$_{\rm
  f}$ flavour symmetries. The impact of releasing these conditions is
investigated and included in the systematic uncertainty.  The coefficients
$\gamma_k$ are fixed to zero for the two non-singlet distributions.  The
exponent $\beta_{\rm g}$, which is not well determined from the data, is fixed
to $3.0225$ \cite{MSTW} and the uncertainty from the introduced bias is included
in the final uncertainty.  This leaves 11 free parameters in the fitted parton
distributions.  The expression for $\chi^2$ of the fit consists of three terms,
\begin{equation}
\chi^2 = \sum\limits_{n=1}^{N_{exp}} \left[ \sum\limits_{i=1}^{N_n^{data}}
  \left(\frac{g_1^{fit} - \mathcal{N}_{_n} g_{1,i}^{data}}{\mathcal{N}_{_n}
    \sigma_{_i}} \right)^2 + \left( \frac{1-\mathcal{N}_{_n}}{\delta
    \mathcal{N}_{_n}} \right)^2\right] + \chi^2_{\rm positivity}\,.
\label{chisq}
\end{equation}
Only statistical uncertainties of the data are taken into account in $\sigma_i$.
The normalisation factors $\mathcal{N}_{_n}$ of each data set $n$ are allowed to
vary taking into account the normalisation uncertainties $\delta \mathcal
N_{_n}$.  If the latter are unavailable, they are estimated as quadratic sums of
the uncertainties of the beam and target polarisations.  The fitted
normalisations are found to be consistent with unity, except for the E155 proton
data where the normalisation is higher, albeit compatible with the value quoted
in Ref.\,\citen{e155p}.

In order to keep the parameters within their physical ranges, the polarised PDFs
are calculated at every iteration of the fit and required to satisfy the
positivity conditions $\vert\Delta q(x)+\Delta \bar{q}(x)\vert \leq
q(x)+\bar{q}(x)$ and $\vert\Delta g(x)\vert\leq g(x)$ at $Q^2 = 1\,(\GeV/c)^2$,
which is accomplished by the $\chi^2_{\rm positivity}$ term in
Eq.\,(\ref{chisq}).  This procedure leads to asymmetric values of the parameter
uncertainties when the fitted value is close to the allowed limit. The
unpolarised PDFs and the corresponding value of the strong coupling constant
$\alpha_s(Q^2)$ are taken from the MSTW parametrisation \cite{MSTW}.  The impact
of the choice of PDFs is evaluated by using the MRST distributions \cite{MRST02}
for comparison.

In order to investigate the sensitivity of the parametrisation of the polarised
PDFs to the functional forms, the fit is performed for several sets of
functional shapes. These shapes do or do not include the $\gamma_{\rm S}$ and
$\gamma_{\rm g}$ parameters of Eq.\,(\ref{eq:param_fk}) and are defined at
reference scales ranging from $1\,(\GeV/c)^2$ to $63\,(\GeV/c)^2$.  It is
observed \cite{andrieux} that mainly two sets of functional shapes are needed to
span almost entirely the range of the possible $\Delta q^{\rm S}(x)$ and $\Delta
g(x)$ distributions allowed by the data.  These two sets of functional forms
yield two extreme solutions for $\Delta g(x)$. For $\gamma_{\rm g}=\gamma_{\rm
  S}=0$ ($\gamma_{\rm g}=0$ and $\gamma_{\rm S} \ne 0$) a negative (positive)
solution for $\Delta g(x)$ is obtained. Both solutions are parametrised at
$Q_0^2 = 1\,(\GeV/c)^2$ and lead to similar values of the reduced $\chi^2$ of
the fits of about $1.05/$d.o.f. Changes in the fit result that originate from
using other (converging) functional forms are included in the systematic
uncertainty.

\begin{figure}
\includegraphics[width=.99\textwidth]{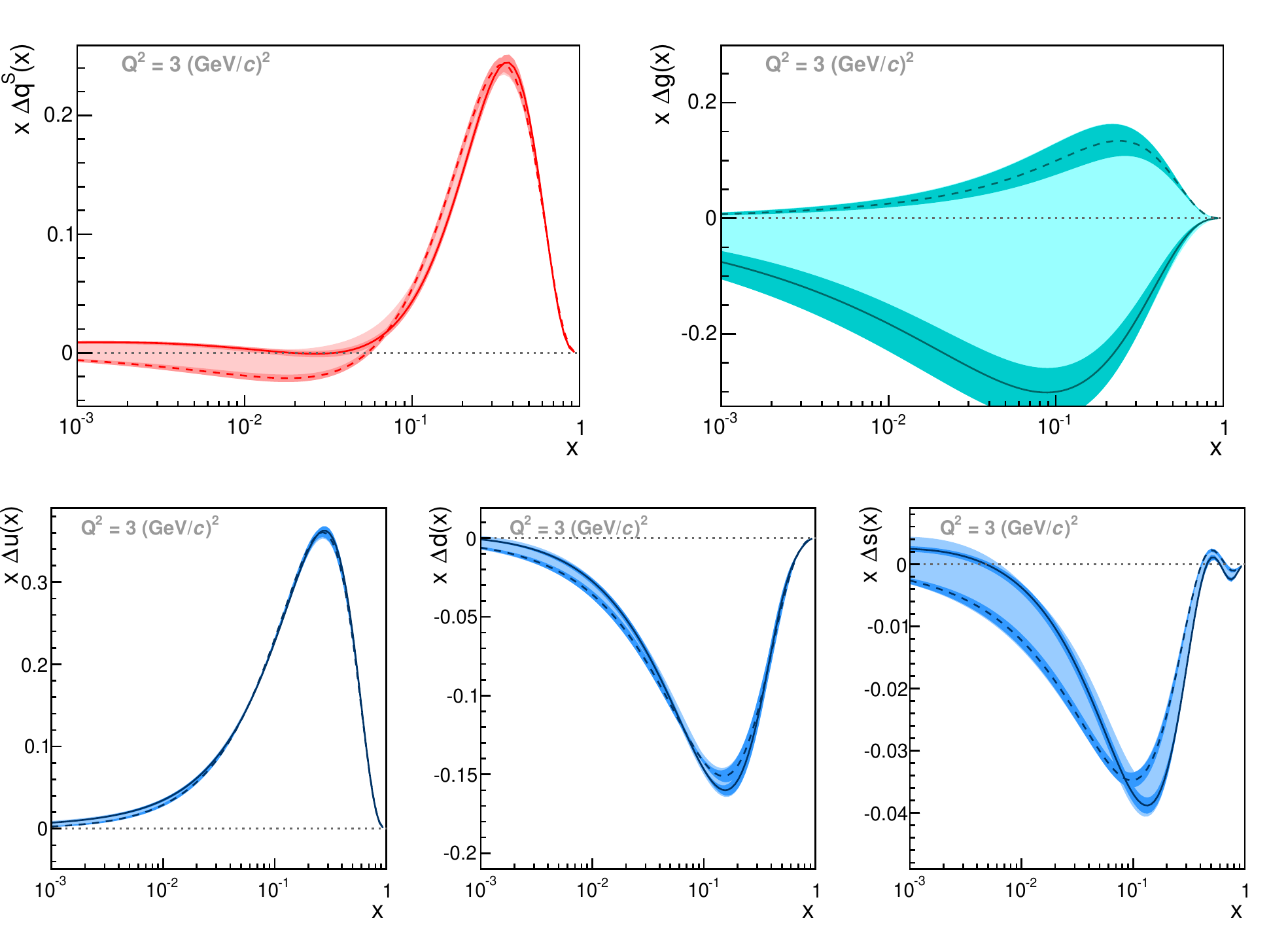}
\caption{Results of the QCD fits to $g_1$ world data at $Q^2 =3\,(\GeV/c)^2$ for
  the two sets of functional shapes as discussed in the text.  Top: singlet
  $x\Delta q^{\rm S}(x)$ and gluon distribution $x\Delta g(x)$.  Bottom:
  distributions of $x\,[\Delta q(x) + \Delta\bar{q}(x)]$ for different flavours
  ($u$, $d$ and $s$). Continuous lines correspond to the fit with $\gamma_{\rm
    S}=0$, long dashed lines to the one with $\gamma_{\rm S}\ne0$. The dark
  bands represent the statistical uncertainties, only. The light bands, which
  overlay the dark ones, represent the systematic uncertainties.}
\label{fig:QCDfitgluons}
\end{figure}

The obtained distributions are presented in Fig.\,\ref{fig:QCDfitgluons}.  The
dark error bands seen in this figure stem from generating several sets of $g_1$
pseudo-data, which are obtained by randomising the measured $g_1$ values using
their statistical uncertainties according to a normal distribution. This
corresponds to a one-standard-deviation accuracy of the extracted parton
distributions.  A thorough analysis of systematic uncertainties of the fitting
procedure is performed. The most important source is the freedom in the choice
of the functional forms for $\Delta q^{\rm S}(x)$ and $\Delta g(x)$.  Further
uncertainties arise from the uncertainty in the value of $\alpha_s(Q^2)$ and
from effects of SU(2)$_f$ and SU(3)$_f$ symmetry breaking.  The systematic
uncertainties are represented by the light bands overlaying the dark ones in
Fig.\,\ref{fig:QCDfitgluons}.  For both sets of functional forms discussed
above, $\Delta s(x)$ stays negative. It is different from zero for $x \gtrsim
0.001$ as are $\Delta d(x)$ and $\Delta u(x)$. The singlet distribution $\Delta
q^{\rm S}(x)$ is compatible with zero for $x \lesssim 0.07$.

The inclusion of systematic uncertainties in the fit leads to much larger
spreads in the first moments as compared to those obtained by only propagating
statistical uncertainties (see Table~\ref{tab:QCDintegral}).  In this table,
$\Delta\Sigma$ denotes the first moment of the singlet distribution. Note that
the first moments of $\Delta u + \Delta\bar u$, $\Delta d + \Delta\bar d$ and
$\Delta s + \Delta\bar s$ are not independent, since the first moments of the
non-singlet distributions are fixed by the decay constants F and D at every
value of $Q^2$.  The large uncertainty in $\Delta g(x)$, which is mainly due to
the freedom in the choice of its functional form, does however not allow to
determine the first moment of $\Delta g(x)$ from the available inclusive data
only.  { \begin{sidewaystable}[htbp]
  \caption{List of experimental data sets used in this analysis.  For each set
    the number of points, the $\chi^2$ contribution and the fitted normalisation
    factor is given for the two functional shapes discussed in the text, which
    lead to either a positive or a negative function $\Delta g(x)$.}
  \label{tab:DataSetInput}
\footnotesize
 \centering
  \begin{tabular}{|l|c|c|c|c|c|c|}\hline
    \multicolumn{1}{|c|}{\multirow{2}{*}{Experiment}}&Function & Number & \multicolumn{2}{|c|}{$\chi^2$} & \multicolumn{2}{|c|}{Normalisation}\\\cline{4-7}
		        	&extracted   			& of points 	& $\Delta g(x)>0$	& $\Delta g(x)<0$	& $\Delta g(x)>0$	& $\Delta g(x) <0$		\\\hline
	EMC \cite{emc}					&  $A_1^{\rm p}$		& 10			& $\hphantom{0}5.2$	& $\hphantom{0}4.7$	& $1.03 \pm 0.07$	& $1.02 \pm 0.07$	\\
	E142 \cite{e142}				& $A_1^{\rm n}$		& \hphantom{0}6	& $\hphantom{0}1.1$	& $\hphantom{0}1.1$	& $1.01 \pm 0.07$	& $0.99 \pm 0.07$	\\
	E143 \cite{e143}				& $g_1^{\rm d}/F_1^{\rm d}$	& 54			& $           61.4$	& $           59.0$	& $0.99 \pm 0.04$	& $1.01 \pm 0.04$	\\
	E143 \cite{e143}				& $g_1^{\rm p}/F_1^{\rm p}$	& 54			& $           47.4$	& $           49.1$	& $1.05 \pm 0.02$	& $1.08 \pm 0.02$	\\
	E154 \cite{e154}				& $A_1^{\rm n}$		& 11			& $\hphantom{0}5.9$	& $\hphantom{0}7.4$	& $1.06 \pm 0.04$	& $1.07 \pm 0.04$	\\
	E155 \cite{e155_d}				& $g_1^{\rm d}/F_1^{\rm d}$	& 22			& $           18.8$	& $           18.0$	& $1.00 \pm 0.04$	& $1.00 \pm 0.04$	\\
	E155 \cite{e155p}				& $g_1^{\rm p}/F_1^{\rm p}$	& 21			& $           50.0$	& $           49.7$	& $1.16 \pm 0.02$	& $1.16 \pm 0.02$	\\
	SMC \cite{smc}					& $A_1^{\rm p}$		& 59			& $           55.4$	& $           55.4$	& $1.02 \pm 0.03$	& $1.01 \pm 0.03$	\\
	SMC \cite{smc}					& $A_1^{\rm d}$		& 65			& $           59.3$	& $           61.5$	& $1.00 \pm 0.04$	& $1.00 \pm 0.04$	\\
	HERMES \cite{HERMES}			& $A_1^{\rm d}$		& 24			& $           28.1$	& $           27.0$	& $0.98 \pm 0.04$	& $1.01 \pm 0.04$	\\
	HERMES \cite{HERMES}			&  $A_1^{\rm p}$		& 24			& $           14.0$	& $           16.2$	& $1.08 \pm 0.03$	& $1.10 \pm 0.03$	\\
	HERMES \cite{hrm1}				& $A_1^{\rm n}$		& \hphantom{0}7	& $\hphantom{0}1.6$	& $\hphantom{0}1.2$	& $1.01 \pm 0.07$	& $1.00 \pm 0.07$	\\
	COMPASS 160 GeV \cite{g1d2006}		& $g_1^{\rm d}$		& 43			& $           33.1$	& $           37.7$	& $0.97 \pm 0.05$	& $0.95 \pm 0.05$	\\
	COMPASS $160\,\GeV$  \cite{g1p2010}		& $A_1^{\rm p}$		& 44			& $           50.8$	& $           49.1$	& $1.00 \pm 0.03$	& $0.99 \pm 0.03$	\\
	COMPASS $200\,\GeV$  (this work)			& $A_1^{\rm p}$		& 51			& $           43.6$	& $           43.2$	& $1.03 \pm 0.03$	& $1.02 \pm 0.03$	\\

    \hline
\end{tabular}
\end{sidewaystable}
}
\begin{table}[htbp]
  \caption{Value ranges of first moments of quark distributions, as obtained
    from the QCD fit when taking into account both statistical and systematic
    uncertainties, as detailed in the text.}
  \label{tab:QCDintegral}
\centering
  \begin{tabular}{|c|c|}\hline
    First moment & {Value range at $Q^2 = 3\,(\GeV/c)^2$}  \\\hline
    $\Delta \Sigma$ & [ \hphantom{$-$}0.26 , \hphantom{$-$}0.36 ] \\
    $\Delta u + \Delta \bar{u}$& [ \hphantom{$-$}0.82 , \hphantom{$-$}0.85 ]\\
    $\Delta d + \Delta \bar{d}$& [ $-$0.45 , $-$0.42 ]\\
    $\Delta s + \Delta \bar{s}$& [ $-$0.11 , $-$0.08 ]\\\hline
  \end{tabular}
\end{table}

\begin{figure}[htbp]
\includegraphics[width=.49\textwidth]{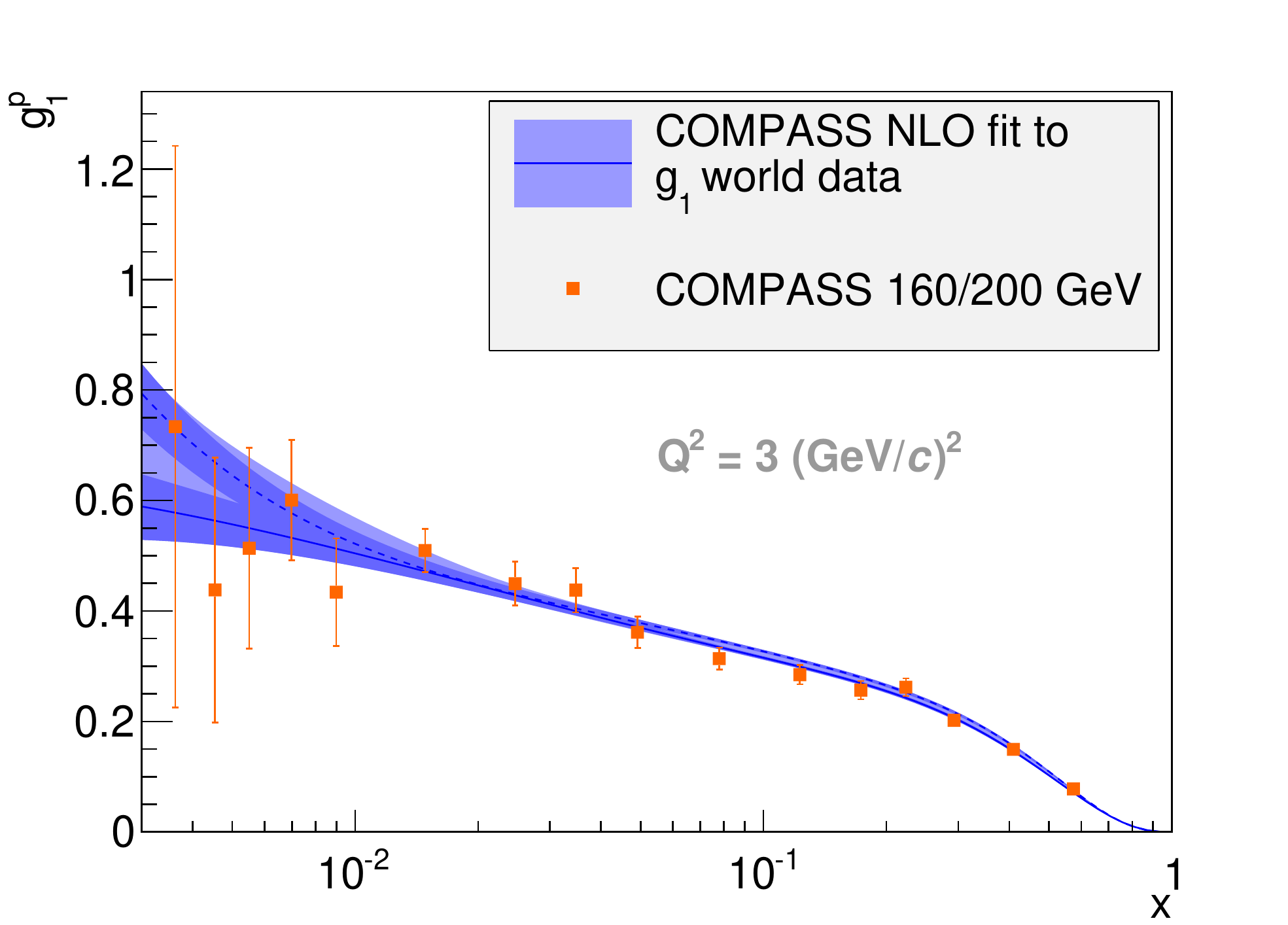}
\includegraphics[width=.49\textwidth]{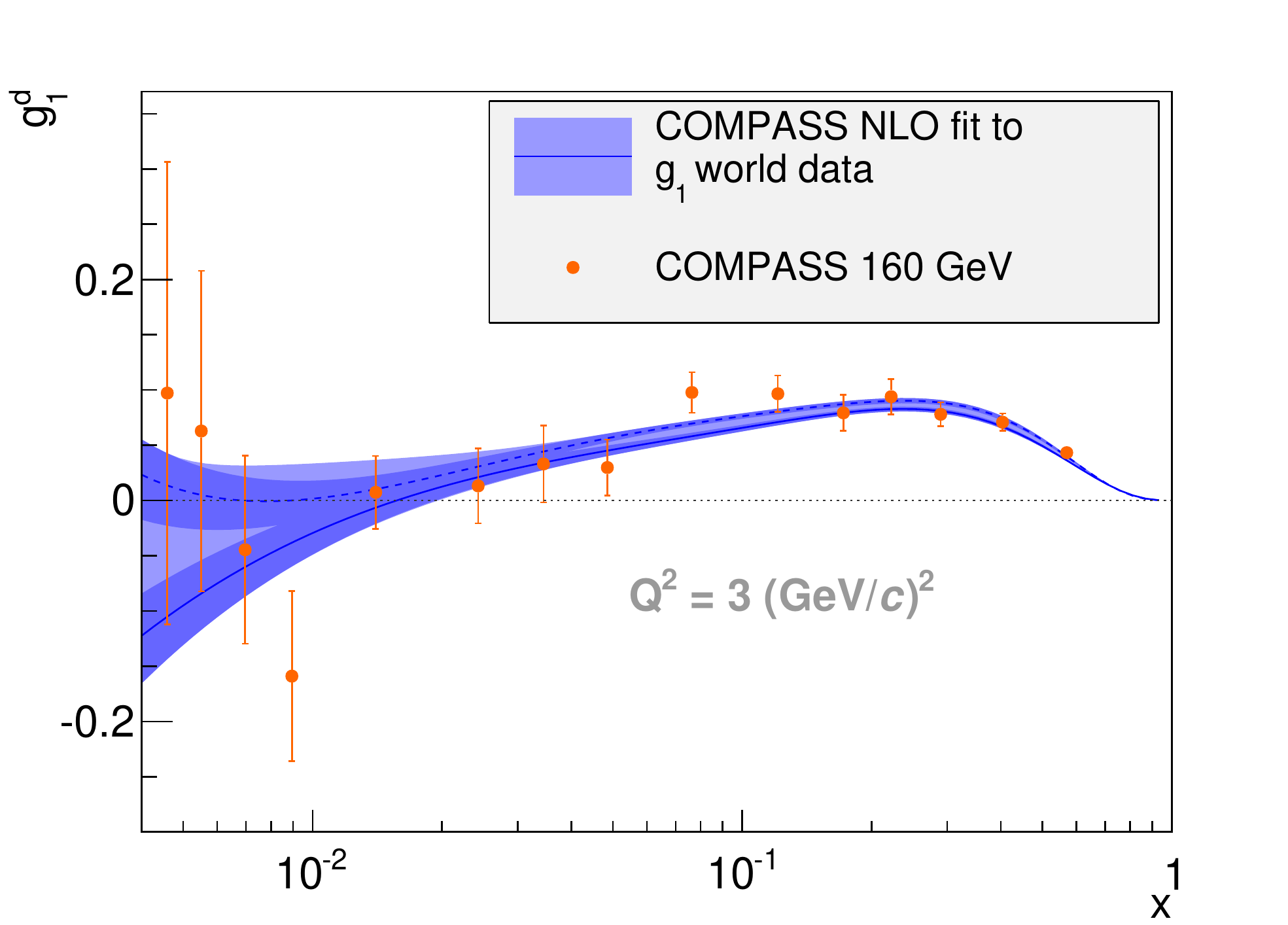}
\caption{Results of the QCD fits to $g_1^{\rm p}$ (left) and $g_1^{\rm d}$
  (right) world data at $Q^2 =3\,(\GeV/c)^2$ as functions of $x$. The curves
  correspond to the two sets of functional shapes as discussed in the text.  The
  dark bands represent the statistical uncertainties associated with each curve
  and the light bands, which overlay the dark ones, represent the systematic
  uncertainties.}
\label{fig:g1p_g1d_final}
\end{figure}

The fitted $g_1^{\rm p}$ and $g_1^{\rm d}$ distributions at $Q^2 = 3~(\GeV/c)^2$
are shown in Fig.\,\ref{fig:g1p_g1d_final} together with the data evolved to the
same scale. The two curves correspond to the two extreme functional forms
discussed above, which lead to either a positive or a negative $\Delta
g(x)$. The dark bands represent the statistical uncertainties associated with
each curve and the light bands represent the total systematic and statistical
uncertainties added in quadrature. The values for $g_1^{\rm p}$ are positive in
the whole measured region down to $x=0.0025$, while $g_1^{\rm d}$ is consistent
with zero at low $x$.

\section{First moments of $g_1$ from COMPASS data and Bjorken Sum Rule}
The new data on $g_1^{\rm p}$ together with the new QCD fit allow a more precise
determination of the first moments $\Gamma_1(Q^2)=\int_0^1 g_1(x,Q^2) \text{d}x$
of the proton, neutron and non-singlet spin structure functions using COMPASS
data only.  The latter one is defined as
\begin{equation}
g_1^{\rm NS}(x,Q^2) = g_1^{\rm p}(x,Q^2) - g_1^{\rm n}(x,Q^2) =\,2\,[g_1^{\rm
    p}(x,Q^2) - g_1^{\rm N}(x,Q^2)]\,.
\label{eq:nonsinglet}
\end{equation}
 The integral $\Gamma_1^{\rm NS}(Q^2)$ at a given value of $Q^2$ is connected to
 the ratio $g_{\rm A}/g_{\rm V}$ of the axial and vector coupling constants via
 the fundamental Bjorken sum rule
\begin{equation}
  \Gamma_1^{\rm NS}(Q^2) = \int_0^1 g_1^{\rm NS}(x,Q^2) \text{d}x = \frac{1}{6}
  \Bigl| \frac{g_{\rm A}}{g_{\rm V}} \Bigr| C_1^{\rm NS}(Q^2)\,,
  \label{Eq:BSR}
\end{equation}
where $C_1^{\rm NS}(Q^2)$ is the non-singlet coefficient function that is
known~\cite{Larin} up to the third order in $\alpha_s(Q^2)$ in perturbative QCD.

Due to small differences in the kinematics of the data sets, all points of the
three \mbox{COMPASS} $g_1$ data sets (Table~\ref{tab:DataSetInput}) are evolved
to the $Q^2$ value of the $160\,\GeV$ proton data.  A weighted average of the
$160\,\GeV$ and $200\,\GeV$ proton data is performed and the points at different
values of $Q^2$ and the same value of $x$ are merged.

For the determination of $\Gamma_1^{\rm p}$ and $\Gamma_1^{\rm d}$, the values
of $g_1^{\rm p}$ and $g_1^{\rm d}$ are evolved to $Q^2=3\,(\GeV/c)^2$ and the
integrals are calculated in the measured ranges of $x$. In order to obtain the
full moments, the QCD fit is used to evaluate the extrapolation to $x=1$ and
$x=0$ (see Table~\ref{tab:contributions}).  The moment $\Gamma_1^{\rm n}$ is
calculated using $g_1^{\rm n}=2g_1^{\rm N} - g_1^{\rm p}$.  The systematic
uncertainties of the moments include the uncertainties of $P_{\rm B}$, $P_{\rm
  T}$, $f$ and $D$.  In addition, the uncertainties from the QCD evolution and
those from the extrapolation are obtained using the uncertainties given in
Section~\ref{sec:QCD}. The full moments are given in Table\,\ref{tab:moments}.
Note that also $\Gamma_1^{\rm N}$ is updated compared to Ref.\,\cite{g1d2006}
using the new QCD fit.

\begin{table}
		\centering
	\caption{Contribution to the first moments of $g_1$ at $Q^2
          =3\,(\GeV/c)^2$ with statistical uncertainties from the COMPASS data.
          Limits in parentheses are applied for the calculation of
          $\Gamma_1^{\rm N}$. The uncertainties of the extrapolations are
          negligible.}
	\label{tab:contributions}
		\begin{tabular}{|rcl|c|c|}
			\hline
			\multicolumn{3}{|c|}{$x$ range}                 	& $\Gamma_1^{\rm p}$ & $\Gamma_1^{\rm N}$\\
			\hline
			$0$  	        	&$\!\!\!-\!\!\!$& $0.0025~(0.004)$ & $0.002$            & $0.000$  \\
			$0.0025~(0.004)$	&$\!\!\!-\!\!\!$& $0.7$		& $0.134 \pm 0.003$  & $0.047 \pm 0.003$ \\ 
			$0.7$  			&$\!\!\!-\!\!\!$& $1.0$         & $0.003$	     & $0.001$  \\
			\hline
		\end{tabular}
\end{table}

\begin{table}
	\centering
	\caption{First moments of $g_1$ at $Q^2 =3\,(\GeV/c)^2$ using COMPASS data only.}
	\label{tab:moments}
\begin{tabular}{|c||cccc|}
\hline
		& $\Gamma_1$			& $\delta \Gamma_1^{\rm stat}$	& $\delta \Gamma_1^{\rm syst}$	& $\delta \Gamma_1^{\rm evol}$	\\
\hline
Proton	& $\phantom{-}0.139$	& $\pm 0.003$		& $\pm 0.009$		& $\pm 0.005$		\\
Nucleon	& $\phantom{-}0.049$	& $\pm 0.003$		& $\pm 0.004$		& $\pm 0.004$		\\
Neutron	& $-0.041$ 				& $\pm 0.006$		& $\pm 0.011$		& $\pm 0.005$		\\
\hline
\end{tabular}
\end{table}

\begin{figure}[htbp]
	\begin{minipage}[t]{0.48\textwidth}
		\begin{center}
			\includegraphics[width=\textwidth]{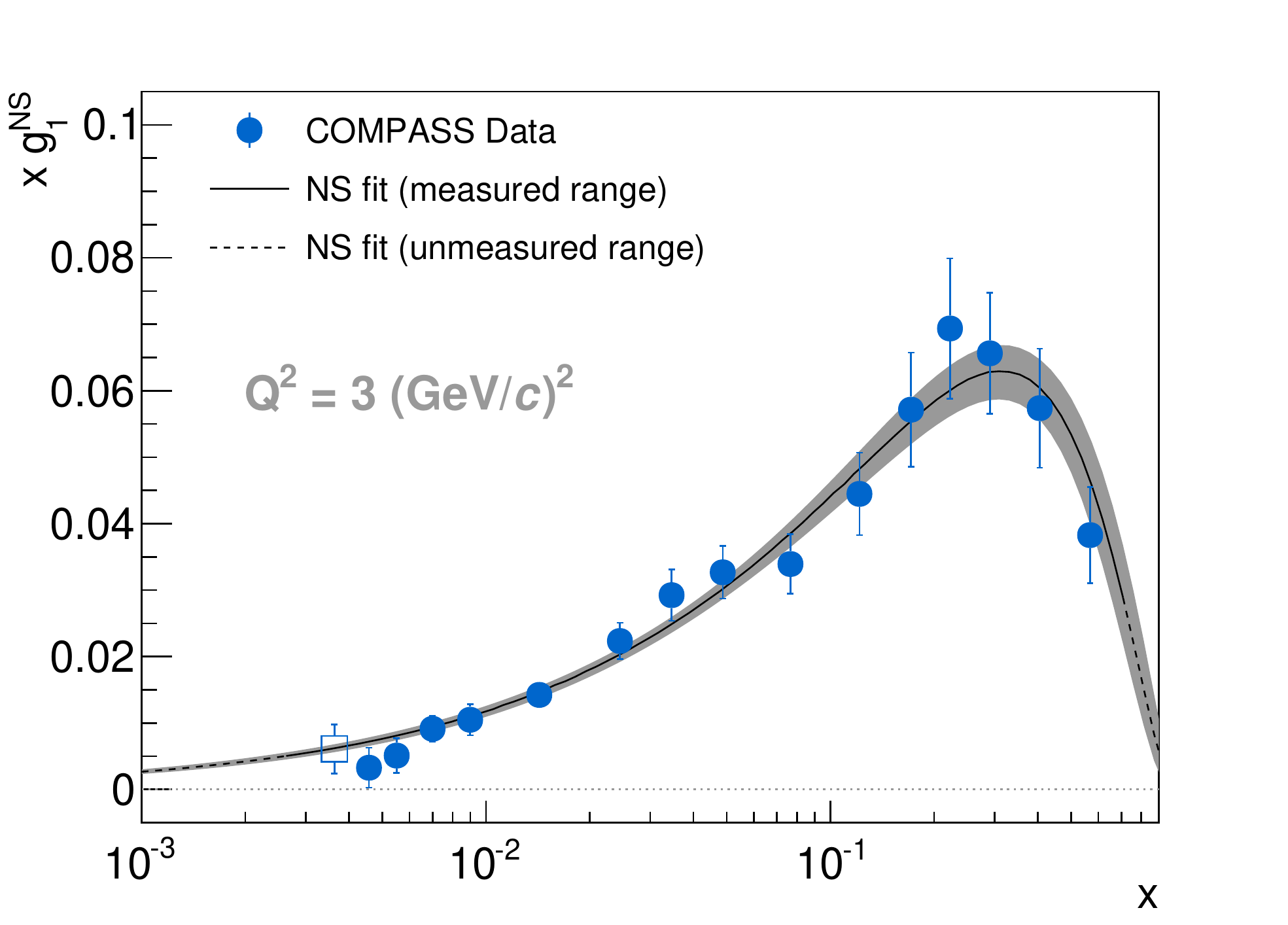}
		\end{center}
		\caption{
			Values of $xg_1^{\rm NS}(x)$ at $Q^2=3\,(\GeV/c)^2$
                        compared to the non-singlet NLO QCD fit using COMPASS
                        data only. The errors bars are statistical.  The open
                        square at lowest $x$ is obtained with $g_1^{\rm d}$
                        taken from the NLO QCD fit.}
		\label{fig:g1NS}
	\end{minipage}
	\hfill
	\begin{minipage}[t]{0.48\textwidth}
		\begin{center}
			\includegraphics[width=\textwidth]{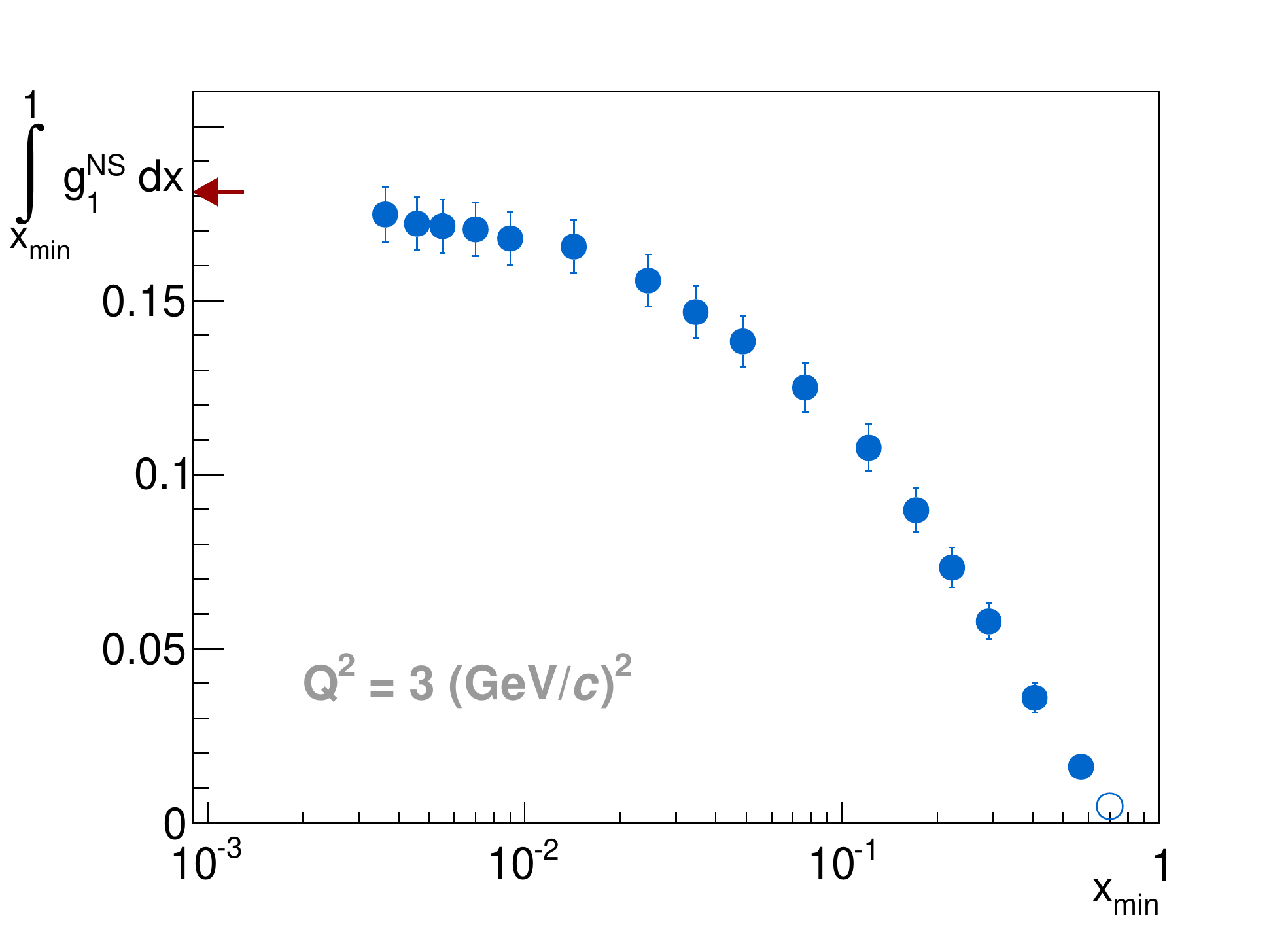}
		\end{center}
		\caption{
			Values of $\int_{x_{\rm min}}^1 g_1^{\rm NS} \text{d}x$ as a
                        function of $x_{\rm min}$. The open circle at $x=0.7$ is
                        obtained from the fit.  The arrow on the left side shows
                        the value for the full range, $0 \le x \le 1$.}
		\label{fig:intNS}
	\end{minipage}
\end{figure}

For the evaluation of the Bjorken sum rule, the procedure is slightly
modified. Before evolving from the measured $Q^2$ to $Q^2=3\,(\GeV/c)^2$,
$g_1^{\rm NS}$ is calculated from the proton and deuteron $g_1$ data.  Since
there is no measured COMPASS value of $g_1^{\rm d}$ corresponding to the new
$g_1^{\rm p}$ point at $x=0.0036$, the value of $g_1^{\rm d}$ from the NLO QCD
fit is used in this case. The fit of $g_1^{\rm NS}$ is performed with the same
program as discussed in the previous section but fitting only the non-singlet
distribution $\Delta q_3(x,Q^2)$.  The parameters of this fit are given in
Table~\ref{tab:NS_fit} and a comparison of the fitted distribution with the data
points is shown in Fig.\,\ref{fig:g1NS}. The error band is obtained with the
same method as described in the previous section.

\begin{table}
	\begin{minipage}[!h]{0.44\textwidth}
		\centering
		\caption{Results of the fit of $\Delta q_3(x)$ at $Q_0^2= 1\,(\GeV/c)^2$. \vspace*{1.3cm}}
		\label{tab:NS_fit}
		\begin{tabular}{|c|rcl|}
			\hline
			Param. & \multicolumn{3}{|c|}{Value}  \\
			\hline
			$\eta_3$  & $1.24$ &$\pm$ &$0.06$ \\
			$\alpha_3$& $-0.11$& $\pm$ &$0.08$ \\
			$\beta_3$ & $ 2.2$& $_{-}^{+}$ & $_{0.4}^{0.5}$ \\ 
			\hline
			$\chi^2/$NDF &\multicolumn{3}{|c|}{ $7.9/13$}  \\
			\hline
		\end{tabular}
	\end{minipage}
	\hfill
	\begin{minipage}[!h]{0.52\textwidth}
		\centering
		\caption{ First moment $\Gamma_1^{\rm NS}$ at $Q^2
                  =3\,(\GeV/c)^2$\, from the COMPASS data with statistical
                  uncertainties. Contributions from the unmeasured regions are
                  estimated from the NLO fit to $g_1^{\rm NS}$. The uncertainty
                  is determined using the error band shown in
                  Fig.\,\ref{fig:g1NS}.}
		\label{tab:NS_integ}
		\begin{tabular}{|rcl|l|}
			\hline
			\multicolumn{3}{|c}{$x$ range}  & \multicolumn{1}{|c|}{$\Gamma_1^{\rm NS}$} \\
			\hline
			$0$  &$\!\!\!-\!\!\!$& $0.0025$ &  $0.006\pm 0.001$          \\
			$0.0025$ &$\!\!\!-\!\!\!$& $0.7$    &  $0.170\pm 0.008$  \\ 
			$0.7$  &$\!\!\!-\!\!\!$& $1.0$    &  $0.005\pm 0.002$          \\
			\hline
			$0$  &$\!\!\!-\!\!\!$& $1$      &  $0.181\pm0.008$   \\
			\hline
		\end{tabular}
	\end{minipage}
\end{table}

The integral of $g_1^{\rm NS}$ in the measured range of $0.0025 < x < 0.7$ is
calculated using the data points. The contribution from the unmeasured region is
extracted again from the fit. The various contributions are listed in
Table~\ref{tab:NS_integ} and the dependence of $\Gamma_1^{\rm NS}$ on the lower
limit of the integral is shown in Fig.\,\ref{fig:intNS}.  The contribution of
the measured $x$ range to the integral corresponds to $93.8\%$ of the full first
moment, while the extrapolation to 0 and 1 amounts to $3.6\%$ and $2.6\%$,
respectively.  Compared to the previous result~\cite{g1p2010}, the contribution
of the extrapolation to $x=0$ is now by about one third smaller than before due
to the larger $x$ range of the present data. The value of the integral for the
full $x$ range is
\begin{equation}
\Gamma_1^{\rm NS} = 0.181\pm 0.008~(\text{stat.}) \pm 0.014~(\text{syst.})~.
\end{equation}

The uncertainty of $\Gamma_1^{\rm NS}$ is dominated by the systematic
uncertainties. The largest contribution stems from the uncertainty of the beam
polarisation (5\%); other contributions originate from uncertainties in the
combined proton data, i.e. those of target polarisation, dilution factor and
depolarisation factor. The uncertainties in the deuteron data have a smaller
impact as the first moment of $g_1^{\rm d}$ is smaller than that of the
proton. The uncertainty due to the evolution to a common $Q^2$ is found to be
negligible when varying $Q_0^2$ between $1\,(\GeV/c)^2$ and
$10\,(\GeV/c)^2$. The overall result agrees well with our earlier result
$\Gamma_1^{\rm NS}=0.190 \pm 0.009 \pm 0.015$ in \cite{g1p2010}.

The result for $\Gamma_1^{\rm NS}$ is used to evaluate the Bjorken sum rule with
Eq.\,(\ref{Eq:BSR}). Using the coefficient function $C_1^{\rm NS}(Q^2)$ at NLO
and $\alpha_s= 0.337$ at $Q^2 = 3 \,(\GeV/c)^2$, one obtains

\begin{equation}
|g_{\rm A} / g_{\rm V}| = 1.22 \pm 0.05~({\rm stat.}) \pm 0.10~(\text{syst.}) .
\end{equation}

The comparison of the value of $| g_{\rm A}/g_{\rm V} |$ from the present
analysis and the one obtained from neutron $\beta$ decay, $|g_{\rm A}/g_{\rm V}|
= 1.2701\pm 0.002$\cite{PDG12}, provides a validation of the Bjorken sum rule
with an accuracy of $9\%$.  Note that the contribution of $\Delta g$ cancels in
Eq.\,(\ref{eq:nonsinglet}) and hence does not enter the Bjorken sum.
Higher-order perturbative corrections are expected to increase slightly the
result.  By using the coefficient function $C_1^{\rm NS}$ at NNLO instead of
NLO, $| g_{\rm A}/g_{\rm V} |$ is found to be 1.25, closer to values stemming
from the neutron weak decay.

\section{Conclusions}
The COMPASS Collaboration performed new measurements of the longitudinal double
spin asymmetry $A_1^{\rm p}(x,Q^2)$ and the longitudinal spin structure function
$g_1^{\rm p}(x,Q^2)$ of the proton in the range $0.0025 < x < 0.7$ and in the
DIS region, $1 < Q^2 < 190\,(\GeV/c)^2$, thus extending the previously covered
kinematic range \cite{g1p2010} towards large values of $Q^2$ and small values of
$x$. The new data improve the statistical precision of $g_1^{\rm p}(x)$ by about
a factor of two for $x \lesssim $ 0.02.

The world data for $g_1^{\rm p}$, $g_1^{\rm d}$ and $g_1^{\rm n}$ were used to
perform a NLO QCD analysis, including a detailed investigation of systematic
effects. This analysis thus updates and supersedes the previous COMPASS QCD
analysis \cite{g1d2006}.  It was found that the contribution of quarks to the
nucleon spin, $\Delta \Sigma$, lies in the interval 0.26 and 0.36 at $Q^2
=3\,(\GeV/c)^2$, where the interval limits reflect mainly the large uncertainty
in the determination of the gluon contribution.

When combined with the previously published results on the deuteron
\cite{g1d2006}, the new $g_1^{\rm p}$ data provide a new determination of the
non-singlet spin structure function $g_1^{\rm NS}$ and a new evaluation of the
Bjorken sum rule, which is validated to an accuracy of about 9~\%.

\section*{Acknowledgements}
We gratefully acknowledge the support of the CERN management and staff and the
skill and effort of the technicians of our collaborating institutes.  This work
was made possible by the financial support of our funding agencies.  

\appendix 
{\setlength{\textheight}{53\baselineskip}
\section{Asymmetry Results}
\label{sec:A1p}
\begin{table}
\caption{Values of $A_1^{\rm p}$ and $g_1^{\rm p}$ as a function of $x$ at 
the measured values of $Q^2$. The first uncertainty is statistical, the second one systematic.}
\label{tab:a_g_values}
\footnotesize
\begin{tabular}{|rcl|c|c|c|c|}
\hline
\multicolumn{3}{|c|}{$x$ range} 	& $\langle x\rangle$ & $Q^2\,((\GeV/c)^2)$ & $A_1^{\rm p}$ 	& $g_1^{\rm p}$ \\\hline
$0.0025$&$\!\!\!-\!\!\!$& $0.004$	& $0.0035$	& $1.03$	& $0.059 \pm 0.029 \pm 0.014$	& $1.79  \pm 0.87 \pm 0.45$\\
		&				&			& $0.0036$	& $1.10$	& $-0.004 \pm 0.027 \pm 0.012$	& $-0.12 \pm 0.81 \pm 0.37$\\
		&				&			& $0.0038$	& $1.22$	& $0.002 \pm 0.032 \pm 0.012$	& $0.05  \pm 0.98 \pm 0.37$\\\hline
$0.004$&$\!\!\!-\!\!\!$& $0.005$	& $0.0044$	& $1.07$	& $0.006 \pm 0.021 \pm 0.008$	& $0.15  \pm 0.50 \pm 0.19$\\
		&				&			& $0.0045$	& $1.24$	& $0.021 \pm 0.020 \pm 0.008$	& $0.53  \pm 0.51 \pm 0.20$\\
		&				&			& $0.0046$	& $1.44$	& $0.023 \pm 0.022 \pm 0.011$	& $0.60  \pm 0.59 \pm 0.28$\\\hline
$0.005$&$\!\!\!-\!\!\!$& $0.006$	& $0.0055$	& $1.11$	& $0.009 \pm 0.024 \pm 0.011$	& $0.18  \pm 0.46 \pm 0.21$\\
		&				&			& $0.0055$	& $1.36$	& $0.026 \pm 0.020 \pm 0.008$	& $0.56  \pm 0.42 \pm 0.17$\\
		&				&			& $0.0056$	& $1.68$	& $0.022 \pm 0.020 \pm 0.008$	& $0.51  \pm 0.47 \pm 0.18$\\\hline
$0.006$&$\!\!\!-\!\!\!$& $0.008$	& $0.0069$	& $1.14$	& $0.033 \pm 0.020 \pm 0.009$	& $0.50  \pm 0.32 \pm 0.14$\\
		&				&			& $0.0069$	& $1.50$	& $0.041 \pm 0.015 \pm 0.007$	& $0.71  \pm 0.27 \pm 0.12$\\
		&				&			& $0.0071$	& $2.02$	& $0.006 \pm 0.014 \pm 0.007$	& $0.12  \pm 0.27 \pm 0.13$\\\hline
$0.008$&$\!\!\!-\!\!\!$& $0.0010$	& $0.0089$	& $1.17$	& $0.007 \pm 0.027 \pm 0.013$	& $0.08  \pm 0.32 \pm 0.16$\\
		&				&			& $0.0089$	& $1.62$	& $0.029 \pm 0.018 \pm 0.007$	& $0.40  \pm 0.25 \pm 0.10$\\
		&				&			& $0.0090$	& $2.41$	& $0.015 \pm 0.014 \pm 0.006$	& $0.24  \pm 0.23 \pm 0.09$\\\hline
$0.010$&$\!\!\!-\!\!\!$& $0.014$	& $0.0116$	& $1.21$	& $0.044 \pm 0.026 \pm 0.013$	& $0.41  \pm 0.24 \pm 0.12$\\
		&				&			& $0.0117$	& $1.75$	& $0.040 \pm 0.017 \pm 0.011$	& $0.42  \pm 0.18 \pm 0.11$\\
		&				&			& $0.0120$	& $2.92$	& $0.044 \pm 0.011 \pm 0.005$	& $0.56  \pm 0.14 \pm 0.07$\\\hline
$0.014$&$\!\!\!-\!\!\!$& $0.020$	& $0.0164$	& $1.26$	& $0.087 \pm 0.034 \pm 0.015$	& $0.58  \pm 0.22 \pm 0.11$\\
		&				&			& $0.0165$	& $1.92$	& $0.100 \pm 0.020 \pm 0.011$	& $0.77  \pm 0.16 \pm 0.09$\\
		&				&			& $0.0168$	& $3.74$	& $0.063 \pm 0.011 \pm 0.006$	& $0.60  \pm 0.10 \pm 0.06$\\\hline
$0.020$&$\!\!\!-\!\!\!$& $0.030$	& $0.0239$	& $1.55$	& $0.072 \pm 0.030 \pm 0.016$	& $0.36  \pm 0.15 \pm 0.08$\\
		&				&			& $0.0240$	& $2.49$	& $0.079 \pm 0.025 \pm 0.011$	& $0.45  \pm 0.14 \pm 0.07$\\
		&				&			& $0.0246$	& $5.16$	& $0.079 \pm 0.011 \pm 0.008$	& $0.545 \pm 0.077 \pm 0.061$\\\hline
$0.030$&$\!\!\!-\!\!\!$& $0.040$	& $0.0341$	& $2.18$	& $0.103 \pm 0.035 \pm 0.016$	& $0.39  \pm 0.13 \pm 0.06$\\
		&				&			& $0.0343$	& $3.50$	& $0.099 \pm 0.041 \pm 0.018$	& $0.43  \pm 0.18 \pm 0.08$\\
		&				&			& $0.0347$	& $7.07$	& $0.083 \pm 0.015 \pm 0.013$	& $0.421 \pm 0.075 \pm 0.066$\\\hline
$0.040$&$\!\!\!-\!\!\!$& $0.060$	& $0.0473$	& $2.65$	& $0.128 \pm 0.040 \pm 0.023$	& $0.37  \pm 0.12 \pm 0.07$\\
		&				&			& $0.0480$	& $5.00$	& $0.136 \pm 0.026 \pm 0.016$	& $0.449 \pm 0.086 \pm 0.057$\\
		&				&			& $0.0492$	& $10.4$	& $0.103 \pm 0.015 \pm 0.011$	& $0.378 \pm 0.056 \pm 0.043$\\\hline
$0.060$&$\!\!\!-\!\!\!$& $0.100$	& $0.0740$	& $4.91$	& $0.147 \pm 0.031 \pm 0.016$	& $0.308 \pm 0.066 \pm 0.036$\\
		&				&			& $0.0754$	& $10.7$	& $0.203 \pm 0.020 \pm 0.017$	& $0.465 \pm 0.047 \pm 0.043$\\
		&				&			& $0.0800$	& $19.7$	& $0.129 \pm 0.023 \pm 0.021$	& $0.292 \pm 0.052 \pm 0.050$\\\hline
$0.100$&$\!\!\!-\!\!\!$& $0.150$	& $0.119$	& $8.23$	& $0.291 \pm 0.038 \pm 0.024$	& $0.397 \pm 0.051 \pm 0.035$\\
		&				&			& $0.121$	& $17.8$	& $0.263 \pm 0.028 \pm 0.021$	& $0.372 \pm 0.040 \pm 0.031$\\
		&				&			& $0.125$	& $31.7$	& $0.242 \pm 0.034 \pm 0.021$	& $0.337 \pm 0.048 \pm 0.030$\\\hline
$0.150$&$\!\!\!-\!\!\!$& $0.200$	& $0.171$	& $12.9$	& $0.299 \pm 0.045 \pm 0.027$	& $0.279 \pm 0.042 \pm 0.026$\\
		&				&			& $0.172$	& $26.9$	& $0.316 \pm 0.045 \pm 0.036$	& $0.298 \pm 0.042 \pm 0.035$\\
		&				&			& $0.175$	& $43.8$	& $0.344 \pm 0.050 \pm 0.029$	& $0.318 \pm 0.046 \pm 0.028$\\\hline
$0.200$&$\!\!\!-\!\!\!$& $0.250$	& $0.222$	& $16.1$	& $0.405 \pm 0.060 \pm 0.043$	& $0.273 \pm 0.040 \pm 0.030$\\
		&				&			& $0.222$	& $32.1$	& $0.340 \pm 0.066 \pm 0.035$	& $0.227 \pm 0.044 \pm 0.024$\\
		&				&			& $0.224$	& $52.4$	& $0.268 \pm 0.060 \pm 0.045$	& $0.174 \pm 0.039 \pm 0.030$\\\hline
$0.250$&$\!\!\!-\!\!\!$& $0.350$	& $0.289$	& $21.7$	& $0.397 \pm 0.057 \pm 0.035$	& $0.176 \pm 0.025 \pm 0.016$\\
		&				&			& $0.290$	& $42.1$	& $0.374 \pm 0.077 \pm 0.050$	& $0.160 \pm 0.033 \pm 0.022$\\
		&				&			& $0.296$	& $66.3$	& $0.392 \pm 0.062 \pm 0.035$	& $0.159 \pm 0.025 \pm 0.015$\\\hline
$0.350$&$\!\!\!-\!\!\!$& $0.500$	& $0.403$	& $28.4$	& $0.396 \pm 0.086 \pm 0.051$	& $0.085 \pm 0.018 \pm 0.011$\\
		&				&			& $0.405$	& $53.1$	& $0.40  \pm 0.12 \pm 0.06$		& $0.079 \pm 0.024 \pm 0.011$\\
		&				&			& $0.413$	& $85.1$	& $0.631 \pm 0.088 \pm 0.054$	& $0.114 \pm 0.016 \pm 0.010$\\\hline
$0.500$&$\!\!\!-\!\!\!$& $0.700$	& $0.561$	& $29.8$	& $0.42  \pm 0.23 \pm 0.10$		& $0.028 \pm 0.016 \pm 0.006$\\
		&				&			& $0.567$	& $50.4$	& $0.75  \pm 0.23 \pm 0.12$		& $0.044 \pm 0.013 \pm 0.007$\\
		&				&			& $0.575$	& $96.1$	& $0.87  \pm 0.15 \pm 0.09$		& $0.0429 \pm 0.0071 \pm 0.0048$\\\hline
\end{tabular}
\end{table}}

\end{document}

%% file: Authors2015_g1.tex
%
%
\section*{The COMPASS Collaboration}
\label{app:collab}
\renewcommand\labelenumi{\textsuperscript{\theenumi}~}
\renewcommand\theenumi{\arabic{enumi}}
\begin{flushleft}
C.~Adolph\Irefn{erlangen},
R.~Akhunzyanov\Irefn{dubna}, 
M.G.~Alexeev\Irefn{turin_u},
G.D.~Alexeev\Irefn{dubna}, 
A.~Amoroso\Irefnn{turin_u}{turin_i},
V.~Andrieux\Irefn{saclay},
V.~Anosov\Irefn{dubna}, 
A.~Austregesilo\Irefn{munichtu},
C.~Azevedo\Irefn{aveiro},           
B.~Bade{\l}ek\Irefn{warsawu}
% ...
\end{flushleft}
%
%
\begin{Authlist}
\item \Idef{turin_p}{University of Eastern Piedmont, 15100 Alessandria, Italy}
\item \Idef{aveiro}{University of Aveiro, Department of Physics, 3810-193 Aveiro, Portugal} 
\item \Idef{bochum}{Universit\"at Bochum, Institut f\"ur Experimentalphysik, 44780 Bochum, Germany\Arefs{l}\Arefs{s}}
\item \Idef{bonniskp}{Universit\"at Bonn, Helmholtz-Institut f\"ur  Strahlen- und Kernphysik, 53115 Bonn, Germany\Arefs{l}}
\item \Idef{bonnpi}{Universit\"at Bonn, Physikalisches Institut, 53115 Bonn, Germany\Arefs{l}}
\item \Idef{brno}{Institute of Scientific Instruments, AS CR, 61264 Brno, Czech Republic\Arefs{m}}
\item \Idef{calcutta}{Matrivani Institute of Experimental Research \& Education, Calcutta-700 030, India\Arefs{n}}
\item \Idef{dubna}{Joint Institute for Nuclear Research, 141980 Dubna, Moscow region, Russia\Arefs{o}}
\item \Idef{erlangen}{Universit\"at Erlangen--N\"urnberg, Physikalisches Institut, 91054 Erlangen, Germany\Arefs{l}}
\item \Idef{freiburg}{Universit\"at Freiburg, Physikalisches Institut, 79104 Freiburg, Germany\Arefs{l}\Arefs{s}}
\item \Idef{cern}{CERN, 1211 Geneva 23, Switzerland}
\item \Idef{liberec}{Technical University in Liberec, 46117 Liberec, Czech Republic\Arefs{m}}
\item \Idef{lisbon}{LIP, 1000-149 Lisbon, Portugal\Arefs{p}}
\item \Idef{mainz}{Universit\"at Mainz, Institut f\"ur Kernphysik, 55099 Mainz, Germany\Arefs{l}}
\item \Idef{miyazaki}{University of Miyazaki, Miyazaki 889-2192, Japan\Arefs{q}}
\item \Idef{moscowlpi}{Lebedev Physical Institute, 119991 Moscow, Russia}
\item \Idef{munichtu}{Technische Universit\"at M\"unchen, Physik Department, 85748 Garching, Germany\Arefs{l}\Arefs{r}}
\item \Idef{nagoya}{Nagoya University, 464 Nagoya, Japan\Arefs{q}}
\item \Idef{praguecu}{Charles University in Prague, Faculty of Mathematics and Physics, 18000 Prague, Czech Republic\Arefs{m}}
\item \Idef{praguectu}{Czech Technical University in Prague, 16636 Prague, Czech Republic\Arefs{m}}
\item \Idef{protvino}{State Scientific Center Institute for High Energy Physics of National Research Center `Kurchatov Institute', 142281 Protvino, Russia}
\item \Idef{saclay}{CEA IRFU/SPhN Saclay, 91191 Gif-sur-Yvette, France\Arefs{s}}
\item \Idef{taipei}{Academia Sinica, Institute of Physics, Taipei, 11529 Taiwan}
\item \Idef{telaviv}{Tel Aviv University, School of Physics and Astronomy, 69978 Tel Aviv, Israel\Arefs{t}}
\item \Idef{triest_u}{University of Trieste, Department of Physics, 34127 Trieste, Italy}
\item \Idef{triest_i}{Trieste Section of INFN, 34127 Trieste, Italy}
\item \Idef{triest_ictp}{Abdus Salam ICTP, 34151 Trieste, Italy}
\item \Idef{turin_u}{University of Turin, Department of Physics, 10125 Turin, Italy}
\item \Idef{turin_i}{Torino Section of INFN, 10125 Turin, Italy}
\item \Idef{illinois}{University of Illinois at Urbana-Champaign, Department of Physics, Urbana, IL 61801-3080, U.S.A.}   
\item \Idef{warsaw}{National Centre for Nuclear Research, 00-681 Warsaw, Poland\Arefs{u} }
\item \Idef{warsawu}{University of Warsaw, Faculty of Physics, 02-093 Warsaw, Poland\Arefs{u} }
\item \Idef{warsawtu}{Warsaw University of Technology, Institute of Radioelectronics, 00-665 Warsaw, Poland\Arefs{u} }
\item \Idef{yamagata}{Yamagata University, Yamagata, 992-8510 Japan\Arefs{q} }
\end{Authlist}
%
%
\vspace*{-\baselineskip}\renewcommand\theenumi{\alph{enumi}}
\begin{Authlist}
\item \Adef{a0}{Retired from Universit\"at Bielefeld, Fakult\"at f\"ur Physik, 33501 Bielefeld, Germany}
\item \Adef{a}{Also at Instituto Superior T\'ecnico, Universidade de Lisboa, Lisbon, Portugal}
\item \Adef{a1}{Present address: Universit\"at Mainz, Helmholtz-Institut f\"ur Strahlen- und Kernphysik, 55099 Mainz, Germany}
\item \Adef{b}{Also at Department of Physics, Pusan National University, Busan 609-735, Republic of Korea and at Physics Department, Brookhaven National Laboratory, Upton, NY 11973, U.S.A. }
\item \Adef{c}{Supported by the DFG Research Training Group Programme 1102  ``Physics at Hadron Accelerators''}
\item \Adef{d}{Also at Chubu University, Kasugai, Aichi, 487-8501 Japan\Arefs{q}}
\item \Adef{e}{Also at KEK, 1-1 Oho, Tsukuba, Ibaraki, 305-0801 Japan}
\item \Adef{f}{Present address: Universit\"at Bonn, Helmholtz-Institut f\"ur Strahlen- und Kernphysik, 53115 Bonn, Germany}
\item \Adef{g}{Also at Moscow Institute of Physics and Technology, Moscow Region, 141700, Russia}
\item \Adef{h}{Present address: RWTH Aachen University, III. Physikalisches Institut, 52056 Aachen, Germany}
\item \Adef{i}{Present address: Uppsala University, Box 516, SE-75120 Uppsala, Sweden}
\item \Adef{l}{Supported by the German Bundesministerium f\"ur Bildung und Forschung}
\item \Adef{m}{Supported by Czech Republic MEYS Grant LG13031}
\item \Adef{n}{Supported by SAIL (CSR), Govt.\ of India}
\item \Adef{o}{Supported by CERN-RFBR Grant 12-02-91500}
\item \Adef{p}{\raggedright Supported by the Portuguese FCT - Funda\c{c}\~{a}o para a Ci\^{e}ncia e Tecnologia, COMPETE and QREN, Grants CERN/FP/109323/2009, CERN/FP/116376/2010 and CERN/FP/123600/2011}
\item \Adef{q}{Supported by the MEXT and the JSPS under the Grants No.18002006, No.20540299 and No.18540281; Daiko Foundation and Yamada Foundation}
\item \Adef{r}{Supported by the DFG cluster of excellence `Origin and Structure of the Universe' (www.universe-cluster.de)}
\item \Adef{s}{Supported by EU FP7 (HadronPhysics3, Grant Agreement number 283286)}
\item \Adef{t}{Supported by the Israel Science Foundation, founded by the Israel Academy of Sciences and Humanities}
\item \Adef{u}{Supported by the Polish NCN Grant DEC-2011/01/M/ST2/02350}
\item [{\makebox[2mm][l]{\textsuperscript{*}}}] Deceased
\end{Authlist}